\documentclass[aps,nofootinbib,showpacs,twocolumn]{revtex4-1}
\usepackage{graphicx}
\usepackage{dcolumn}
\usepackage{amsmath}
\usepackage{latexsym}
\usepackage{mathrsfs}
\usepackage[usenames,dvipsnames]{color}

\usepackage{hyperref}

\newcommand{\be}{\begin{equation}} 
\newcommand{\ee}{\end{equation}} 
\newcommand{\bea}{\begin{eqnarray}} 
\newcommand{\eea}{\end{eqnarray}} 
\newcommand{\vdt}{\Delta T}
\begin{document}
\title{Hidden superuniversality in systems with continuous variation of critical exponents}
\author{Indranil Mukherjee}
\email{im20rs148@iiserkol.ac.in}
\author{P. K. Mohanty}
\email{pkmohanty@iiserkol.ac.in}
\affiliation {Department of Physical Sciences, Indian Institute of Science Education and Research Kolkata, Mohanpur, 741246 India.}

\begin{abstract}
Renormalization group theory allows continuous variation of critical exponents along a marginal direction (when there is one), keeping the scaling relations invariant. We propose a super universality hypothesis (SUH) suggesting that, up to constant scale factors, the scaling functions along the critical line must be identical to that of the base universality class even when all the critical exponents vary continuously. We demonstrate this in the 
Ashkin Teller (AT) model on a two-dimensional square lattice where two different phase transitions occur across the self-dual critical line:
while magnetic transition obeys the weak-universality hypothesis where exponent ratios remain fixed, the polarization exhibits a continuous variation of all critical exponents. The SUH not only explains both kinds of variations observed in the AT model, it also provides a unified picture of continuous variation of critical exponents observed in several other contexts. 
\end{abstract}\maketitle
\section{Introduction}
Phase transition and critical phenomena \cite{Baxter, Stanley_1971, Hu_2014, Zhu_2020} have been emergent topics of research for several decades. Criticality is associated with two basic features, (a) universality \cite{Griffiths_1970, Stanley_1999}, which states that the associated critical exponents and scaling functions are universal up to symmetries and space dimensionality, and (b) scaling theory \cite{Kadanoff_wsp}, that describes the general properties of the scaling functions and relates different critical exponents. The divergence of correlation length at the critical point of a second-order phase transition ensures that microscopic details of the system have no roles to play - which explains the scale invariant and the universal behavior observed there. However, several experimental systems \cite{Guggenheim, Back, Suzuki} appear to violate this universality hypothesis and exhibit continuous variation of critical exponents w.r.t the system parameters. A clear example is the eight-vertex (8V) model, solved exactly by Baxter \cite{Baxter,Baxter2, Baxter3}, where the critical exponents of the Ferromagnetic transition $\beta$, $\gamma$, $\nu$ change continuously but their ratios $\beta / \nu$ a $\gamma / \nu$, $(2-\alpha)/\nu$ remain invariant. Later, Suzuki \cite{Suzuki} proposed an explanation: the critical exponents should rather be measured w.r.t the correlation length, an emergent length scale of the system, instead of the distance from the critical point which is an {\it external} tuning parameter. This proposal, formally known as the weak universality scenario, explains several experimental features where the continuous variation of exponents is similar to those obtained in 8V model. Now we know that weak universality appears in interacting dimers \cite{Alet}, frustrated spin systems \cite{Queiroz, Jin_Sen}, magnetic hard squares \cite{Pearce}, Blume-Capel models \cite{Malakis}, quantum critical points \cite{Suzuki_Harada}, models of percolation \cite{Andrade_Herrmann, Sahara}, reaction-diffusion systems \cite{Newman}, absorbing phase transition \cite{Noh_Park}, fractal structures \cite{Monceau} etc. 

Most systems that exhibit continuous variation of critical exponents obey weak universality \cite{Suzuki, Guggenheim, Back} where exponents $\beta,\gamma$ and $\nu$ vary but their ratios $\frac\beta\nu, \frac \gamma\nu$ and the field-exponent $\delta$ remain invariant. Another kind of continuous variation where all critical exponents vary except the susceptibility exponent $\gamma$; this is observed in models of fermion mass generation \cite{Kondo} and in the magnetic transition of systems coupled to strain-fields \cite{Puri}. Variation of exponent $\gamma, \nu$
are reported in micellar solutions \cite{Corti, Fisher} but a careful study \cite{Binder4} revealed that it was only a crossover effect. In Ising spin glasses \cite{Bernardi-Campbell} initial study showed continuous variation of exponent $\eta,$ but more recent and accurate results \cite{Vicari2} supported universality with respect to the disorder distribution.

 In all the above examples, one or the other critical exponent remains invariant whereas others vary continuously. 
 Several studies have claimed continuous variation of all the critical exponents in magnetic phase transitions of chemically doped materials, \cite{Butch, Fuchs, Farah, M, T, M11, M40}. In a recent work \cite{pkm_pmandal} it was reported that ferromagnetic transition in Nd-doped single crystal (Sm$_{1-y}$Nd$_y$)$_{0.52}$Sr$_{0.48}$MnO$_3$ all the critical exponents vary continuously with $y$ in the range $(0.5, 1)$. They propose a scaling ansatz that the functional form of the variation must be conditioned to follow the scaling relations. The functional form deduced by this condition could explain the observed variations quite well. They also show that some of the scaling functions are universal along the critical line. The 
 It is, however, not clear if the observed data collapse is only due to 
 the limitations of the data, which is collected in a very small range of temperatures near the critical points. 
 Thus further checks are called for, to assess the invariant nature of the scaling functions along the critical line.

In this article, we propose a super universality hypothesis (SUH): when all or some of the critical exponents vary along a critical line parameterized by a marginal operator what remains universal along the critical line and carries the features of the parent universality are the scaling functions. 
We demonstrate this in the Ashkin Teller (AT) model which is an ideal laboratory to test SUH numerically because in this model the magnetic phase transition follows a weak universality scenario whereas in the electric phase transition, all the critical exponents vary continuously with the interaction parameter; in both cases, we show that the underlying scaling functions are invariant up to multiplicative scale factors.

The article is organized as follows. For completeness, in Sec. II, we define AT model on the square lattice and discuss the magnetic and electric phase transitions, the equation of the critical line, and the critical exponents. A generalized universality hypothesis, namely `super universality hypothesis (SUH)' is proposed in section -III. Here, we obtain several scaling functions of both the transitions in AT model and show that they remain invariant along the critical line.
The results are summarized in section IV.

\section{Exact results and phase transition of The Ashkin Teller model}

The Ashkin Teller (AT) model \cite{AT_1943, Kadanoff_1977, Zisook} is a two-layer Ising system with a marginal four-body interaction between the layers \cite{Fan_Wu}. 
The model on a square lattice can be mapped exactly \cite{Kadanoff_1971} to the well-known eight-vertex (8V) model \cite{Baxter}, where each site of a square lattice is allowed to have non-zero stationary weights for only eight out of sixteen different possible vertices. AT model naturally leads to two different kinds of order parameters - namely magnetic and electric ones. The usual ferromagnetic phase transition clearly belongs to a weak universality scenario whereas the nature of the critical behavior of the electric phase transition has been debated. Recently Kr\ifmmode \check{c}\else \v{c}\fi{}m\'ar et. al. \cite{Roman_Ladislav} have proposed that the electric-phase transition in a symmetric 8V model (and thus the electric transition in AT model) is {\it fully non-universal}.

In AT model on a $L\times L$ square lattice with periodic boundary conditions in both directions, each site ${\bf i}$ of the lattice carries two different Ising spins $\sigma_{\bf i}=\pm$ and $\tau_{\bf i}=\pm.$ The neighbouring spins interact following a Hamiltonian, 
\begin{equation} \label{eq:AT_H}
 H = -J_{\sigma}\sum_{\langle {\bf ij}\rangle} \sigma_{\bf i} \sigma_{\bf j} - J_{\tau} \sum_{\langle {\bf ij}\rangle}\tau_{\bf i} \tau_{\bf j} - \lambda \sum_{\langle {\bf ij}\rangle}\sigma_{\bf i} \sigma_{\bf j} \tau_{\bf i} \tau_{\bf j}. 
\end{equation}
Here ${\bf j}$ is the nearest neighboring site of ${\bf i}$ and $\langle {\bf ij}\rangle$ denotes a pair of nearest-neighbor sites. Here $J_{\sigma}, J_{\tau} >0$ are the strengths of the intra-spin Ferro-magnetic interactions of $\sigma$ and $\tau$ neighboring spins and $\lambda$ represents interactions among them. We consider only the isotropic case $J_{\sigma} = J_{\tau}=J$ where exact results are available from mapping of the model to 8V model \cite{Baxter}.

The Hamiltonian \eqref{eq:AT_H} is invariant under any of the following transformations: $\sigma \rightarrow -\sigma,$ $\tau \rightarrow -\tau,$ or $\sigma \rightarrow \tau.$ Thus one can treat either of $\langle\sigma\rangle= \langle\tau\rangle,$ or $\langle\sigma\tau\rangle$ as the order parameters of the system; the first one characterizes the ferromagnetic to paramagnetic transition where $\langle\sigma\rangle= \langle\tau\rangle \equiv M$ takes a nonzero value whereas $\langle\sigma\tau\rangle\equiv P$ (formally known as the polarization) becomes nonzero during electric phase transition. 
 Unlike the magnetic phase transitions, the electric transitions are less studied in this model \cite{Roman_Ladislav}. A particular question is, whether this transition obeys the universality hypothesis.

 \subsection{The Phase Diagram}
\begin{figure}[t]
\vspace*{.1 cm}
\centering
\includegraphics[height=6cm]{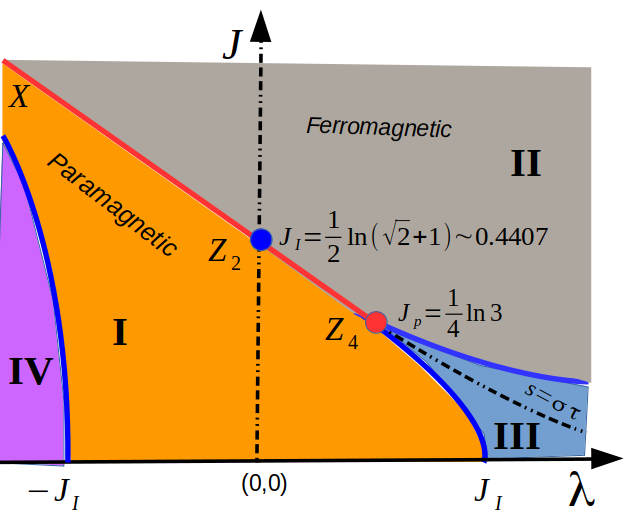}
\caption{(Color online) Phase diagram of AT model on a square lattice for $T=1$ and $J\ge0.$ 
Phase I: Paramagnetic and electrically disordered. Phase II: ferromagnetic and electrically ordered. Phase III: Paramagnetic and electrically ordered.
Phase IV: Paramagnetic and staggered electrical ordered. Staggered magnetic order can also occur for $J<0,$ which is not shown here. 
The critical line $X-Z_4$ separates phase I from phase II; the magnetic and electric phase transition occurs simultaneously as one crosses this self-dual line. Continuous variation of critical exponents along $X-Z_4$ line. The point $Z_2$ where $(\lambda,J)=(0, J_I)$ is the Ising critical point whereas $Z_4$ where $(\lambda, J)=(J_p, J_p)$ is the critical point of four state Potts model.}.
\label{fig:phasediag}
\end{figure}

The phase diagram of the AT model on a square lattice is known exactly from the duality transformations, and from Renormalization-group studies \cite{Wu_Lin,Domany_Riedel}. The phase diagram of the system at temperature $T=1$ and $J\ge0$ is shown in Fig. \ref{fig:phasediag} where $\lambda$ is defined by the duality relation $\sinh (2J) = e^{-2\lambda}.$ 

\noindent{\bf $\lambda = 0$:} For $\lambda = 0$, $\sigma$ and $\tau$ spins are decoupled and Eq. \eqref{eq:AT_H} reduces to two independent Ising systems on a square lattice. Thus, the critical point is $J=J_{I}=\frac{1}{2}\ln(1+\sqrt{2})$ (marked as $Z_{2}$ in Fig. \ref{fig:phasediag}). \\

\noindent { \bf $J=0$:} For $J=0,$ the model reduces to Ising model with a redefined Ising-like spin variable $s_{\bf i} \equiv \sigma_{\bf i} \tau_{\bf i}$ at every site ${\bf i}$ which interact with neighboring spins with interaction strength $\lambda.$ Thus corresponding magnetization $\sum_{\bf i} s_{\bf i}$ can undergo a ferromagnetic transition when $\lambda > \lambda_c = \frac{1}{2}\ln(1+\sqrt{2})$ or an anti-ferromagnetic transition $\lambda <-\lambda_c.$ Note that $\lambda_c$ is same as $J_I.$\\

\noindent{ \bf $\lambda=J$:} For $\lambda=J,$ AT model has a $Z_4$ 
symmetry as Hamiltonian \eqref{eq:AT_H} with $J_\sigma =J_\tau =\lambda$ is invariant under the permutations of the four states ($\{\sigma=\pm\}$, $\{\tau=\pm\}$). Thus, in this case, we have $q=4$ Potts model with the critical point located at $J_p=\lambda=\frac{1}{4}\ln(3)\simeq 0.2746.$ This point is marked as $Z_{4}$ in Fig. \ref{fig:phasediag}.\\

\noindent{\bf $\lambda \rightarrow \infty$:} When $\lambda$ is very large, the terms of $\sigma_{\bf i} \sigma_{\bf j}$ and $\tau_{\bf i} \tau_{\bf j}$ in 
Eq. \eqref{eq:AT_H} must take the same value and their product becomes unity. In this limit, the Hamiltonian reduces to a single site Ising model with coupling $2J.$ Corresponding ferromagnetic Ising critical point is $(J,\lambda ) = (J_c /2, + \infty)$.

AT model has four different phases. Phase I (paramagnetic and electrically disordered): a paramagnetic phase where the couplings are sufficiently weak and none of $M$ and $P$ are ordered, $\langle \sigma \rangle=\langle \tau \rangle=0 = \langle \sigma \tau \rangle.$ Phase II (Ferromagnetic and electrically ordered): the ferromagnetic phase where the couplings are sufficiently strong so that both $M$ and $P$ attain a nonzero value. Phase III and IV (Paramagnetic and electrically ordered): partial ferromagnetic ordering is observed, where $\langle \sigma \tau \rangle$ is ordered ferromagnetically but $\langle \sigma \rangle=\langle \tau \rangle=0$. Phase IV is similar to phase III except that $\langle \sigma \tau \rangle$ is ordered anti-ferromagnetically.

\subsection{Magnetic and electric critical exponents}
The electric and magnetic transitions are characterized by the respective order parameters, the magnetization $M= \sum _{\bf i} \tau_{\bf i}$ $= \sum _{\bf i} \sigma_{\bf i},$ and 
the polarization $P= \sum _{\bf i} \phi_{\bf i},$ where $\phi_{\bf i} = \sigma_{\bf i} \tau_{\bf i}.$ When the temperature of the system is close to critical value $T_c,$
\begin{equation} \label{eq:beta_mag}
 \langle M \rangle \sim \Delta T^{\beta_m} ; ~~ \langle P \rangle\sim \Delta T^{\beta_e}
\end{equation}
where $\Delta T =T_{c}-T$ and $\beta_{m,e}$ are the order parameter  exponents of magnetic, electric transition. The critical exponents $\gamma_{m,e}$ are associated with the susceptibilities 
\begin{equation} %\label{eq:gamma_mag}
 \chi_m = \langle M^{2} \rangle - \langle M \rangle^{2} \sim \Delta T^{-\gamma_m}; \chi_e = \langle P^{2} \rangle - \langle P \rangle^{2} \sim\Delta T^{-\gamma_e}\nonumber
\end{equation}

The correlation functions can be defined as $G_m({\bf r}) =\langle \sigma_{\bf i} \sigma_{\bf i+r} \rangle - \langle M \rangle^{2}$ $=\langle \tau_{\bf i} \tau_{\bf i+r} \rangle - \langle M \rangle^{2}$ and $G_e({\bf r}) =\langle \phi_{\bf i} \phi_{\bf i+r} \rangle- \langle P \rangle^{2}.$ Near the critical point $T=T_c$ 
\begin{equation}
 G_{m,e}(r) \sim \frac{1}{(r/\xi)^{\eta_{m,e}}} e^{(-r/\xi)} \label{eq:corr_fn}
\end{equation}
where $\xi,$ the correlation length, being an emergent length scale of the system does not depend on other details. As one approach the critical point $T_c,$ it diverges as, 
\begin{equation}\label{eq:nu}
 \xi \propto (\Delta T)^{-\nu}.
\end{equation}
Like $\nu,$ the specific heat exponent $\alpha$ of the system, does not carry subscripts $e,m.$ 
\begin{equation}\label{eq:alpha}
 C_v= \langle E^{2} \rangle - \langle E \rangle^{2} \sim (\Delta T)^{-\alpha}.
\end{equation}

In presence of external applied fields $h$ $\tilde h$ that couples to $M$ and $P$ respectively, the 
Hamiltonian is modified.
\bea \label{eq:AT_H_h}
 H &=& -J_{\sigma}\sum_{\langle {\bf ij}\rangle} \sigma_{\bf i} \sigma_{\bf j} - J_{\tau} \sum_{\langle {\bf ij}\rangle}\tau_{\bf i} \tau_{\bf j} - \lambda \sum_{\langle {\bf ij}\rangle}\sigma_{\bf i} \sigma_{\bf j} \tau_{\bf i} \tau_{\bf j}\cr
 &-& h \sum_{\bf i}\left(\sigma_{\bf i}+ \tau_{\bf i}\right) - \tilde h\sum_{\bf i}\sigma_{\bf i} \tau_{\bf i} 
\eea
 Now at $T=T_c,$ $M(h), P(\tilde h)$ scales as, 
\begin{eqnarray}
\label{eq:delta_mag}
 M(h) &\sim& h^{1/\delta_m} {\rm when }~~ \tilde h=0 \cr 
\phi(\tilde h) &\sim& \tilde h^{1/\delta_e} {\rm when } ~~h=0. 
\end{eqnarray}
All the critical exponents are not independent; they are related by scaling relations \cite{Baxter},
\bea \label{eq:scale}
 2-\alpha= d \nu = \gamma \frac{\delta+1} {\delta- 1}= 2\beta+ \gamma= \frac{\gamma d}{2-\eta}
\eea
We expect these scaling relations to be satisfied by the exponents of magnetic and electric transitions, and they indeed do. In fact, the critical exponents of AT model are known from its mapping to 8V model introduced by Baxter \cite{Baxter, Baxter1_1972, Baxter2}, and from re-normalization group arguments \cite{Wu_Lin, Domany_Riedel}. 
\begin{eqnarray}
&&\nu=\frac{2 (\mu-\pi)}{4\mu-3\pi} ~{\rm with} ~ \cos\mu = e^{2\lambda} \sinh(2\lambda); ~ \alpha=2(1-\nu); \cr
&&\beta_e= \frac{2 \nu -1}{4};~
%~ \eta_e= \frac{2 \nu -1}{2 \nu};~
\delta_e=\frac{6 \nu +1}{2 \nu -1}; ~\gamma_e =\frac{1}{2}+\nu
\label{eq:exact_ele_AT}
\\
&&\beta_m=\frac{\nu}{8}; ~
%\eta_m=\frac{1}{4}; ~ 
\delta_m=15; ~\gamma_m =\frac{7\nu}{4}. \label{eq:exact_mag_AT}
\end{eqnarray}
The universal amplitude ratios along the critical line are also known from the equivalence of the model in the scaling limit with the Sine-Gordon quantum field theory \cite{Delfino}.

For magnetic transition, the critical exponents satisfy the 
following relation 
\be
\frac{\beta_m}{\nu} = \beta_m^0,\frac{\gamma_m}{\nu} =  \gamma_m^0, \delta_m =  \delta_m^0, \eta_m= \eta_m^0. \label{eq:WU}
\ee
Here $\{ \beta_m^0=\frac18,$ $ \gamma_m^0= \frac74,  \delta_m^0 = 15,  \eta_m^0=\frac14\}$ are the critical exponents of
the parent universality class (which is the Ising model in $d=2$). This is the well-known weak universality 
 scenario \cite{Suzuki} observed in several experiments \cite{Guggenheim, Back} where $\beta$ and $\gamma$ and $\nu$ varies continuously keeping their ratio $\frac{\beta}{\nu}$ and $\frac{\gamma}{\nu}$ fixed. 

For electric transition, however, all the critical exponents vary with the marginal interaction parameter $\lambda$ and this transition breaks both the universality and weak universality hypothesis; it is not clear if the exponents are related in any way to that of the parent universality class. We propose a generic universality hypothesis (namely SUH) that predicts the functional form of the continuous variations; we show explicitly that the variations observed in electric and magnetic phase transitions of AT model are consistent with SUH.

 \section{The super universality hypothesis} 
The basic assumption of the super universality hypothesis (SUH) is that the continuous variations of critical exponents, whenever they occur, must vary following a functional form that obeys the generic scaling relations (\ref{eq:scale}) and, up to constant scale factors, the scaling functions along the critical line must remain invariant. SUH suggests that, when critical exponents vary continuously, the underlying universal features along the critical line can be read from the universal scaling functions which are invariant up to some constant scale factors.

A generic functional form of variation of the critical exponents can be determined by enforcing their obedience to the scaling relations Eq. (\ref{eq:scale}). In any given dimension $d,$ the variation of $\nu$ determines how $\alpha$ varies but it does not uniquely specify how should $\gamma$ or for that matter $\beta$ vary. In other words, the continuous variation of all the exponents along a critical line (generated by a marginal parameter $\mu$) can be determined by two functions,
\be
\label{eq:var1}
\gamma = \frac{ \gamma_0}{f(\mu)}; ~~ \nu = \frac{\nu_0}{g(\mu)}.
\ee
 Then, the other critical exponents are determined uniquely; 
 in two dimensions ($d=2$), 
\bea\label{eq:Type-III}
&&2-\alpha=\frac{2-\alpha_0}{g(\mu)};
\frac{\delta +1}{\delta -1}=\frac{f(\mu)}{g(\mu)}\frac{ \delta_0 +1}{ \delta_0 -1}\cr 
&& \eta = \frac{g(\mu)} {f(\mu)}  \eta_0 + 2\left( 1- \frac{g(\mu)} {f(\mu)}\right) \cr 
&& \beta = \frac{ \beta_0}{g(\mu)} + \frac{ \gamma_0}{2g(\mu)}\left( 1 - \frac{g(\mu)} {f(\mu)}\right)
\eea
This is the most generic way critical exponents vary.  An obvious spatial case $f(\mu)=1= g(\mu)$ gives universality which is widely observed. Here, the exponents remain invariant along the critical line. The two other special cases which are commonly observed are,
\bea
&{\rm Type-I.} & f(\mu)=g(\mu): \delta ~{\rm and} ~\eta ~{\rm are ~invariant}\cr
& {\rm Type-II.} & f(\mu)=1: \gamma ~ {\rm is~ invariant}
\eea 

Type-I scenario is the well-known weak universality observed both theoretically and experimentally \cite{Suzuki}. In this case, with $g\equiv g(\mu)$ we get, 
\be
\label{eq:Type-I} \nu = \frac{\nu_0}{g}; \gamma = \frac{ \gamma_0}{g}; \beta=\frac{ \beta_0}{g}; \delta=  \delta_0, \eta= \eta_0. \ee
Here, exponents $\delta,\eta,$ and the ratios $\frac \beta \nu$ and $\frac \gamma \nu$ are pinned to the respective values of the base universality class. The ferromagnetic phase transition of AT model is an example of weak universality
with $g(\mu)=\frac{4\mu-3\pi}{2 (\mu-\pi)}.$

Type-II variation is also common. The critical exponents can be written in this case (with $g\equiv g(\mu)$) as, 
\bea
\label{eq:Type-II}
\nu &=& \frac{\nu_0}{g}; \gamma = \gamma_0; \beta=\frac{ \beta_0}{g} + \frac{ \gamma_0}{2 g}(1-g);\cr
\delta&=& \frac{1+ \delta_0 + g ( \delta_0-1)}{ 1+ \delta_0 - g ( \delta_0-1)}; \eta = g \eta_0 + 2 (1-g).
\eea
Some examples of Type-II variations, where $\gamma$ does not change, include mass-generation in QED \cite{Kondo} and magnetic phase transition in the presence of long-ranged strain field \cite{Puri}. In these examples, the varying critical exponents violate both the universality and the weak universality hypothesis.

The most generic scenario is when all the exponents vary.
It is observed in ferromagnetic phase transitions of chemically doped magnetic materials \cite{Farah, M, T, M11, M40, pkm_pmandal }. The electric phase transition in AT model also belongs here and thus, one expects Eq. \eqref{eq:Type-III} to hold. To find the exact form of the functions $f(\mu)= \gamma_e^0/\gamma_e$ and $g(\mu)= \nu_e^0/\nu_e$ we must know the exponents  $\gamma_e^0$ and $\nu_e^0,$ at $\lambda=0.$ The correlation length exponent $\nu_e^0=1$ is the same for both magnetic and electric phase transition.  Since, at $\lambda=0,$ the spin variables $\sigma$ and $\tau$ are independent of each other,  the polarization $P=\langle \sigma \tau \rangle$ must vary as $P= \langle M_\sigma \rangle \langle M_\tau\rangle\sim(\Delta T)^{1/4},$ as $\langle M_{\sigma,\tau} \rangle \sim (\Delta T)^{1/8}.$
 Corresponding variance is then $\chi_e = \langle M_\sigma^2 \rangle \langle M_\tau^2 \rangle - \langle M_\sigma\rangle^2\langle M_\tau\rangle^2.$ Since $\langle M_{\sigma,\tau}^2 \rangle = \chi_{\sigma,\tau} + 2 \langle M_{\sigma,\tau} \rangle^2$ and $\chi_{\sigma,\tau}\sim (\Delta T)^{-7/4},$ we get the dominant variation of $\chi_e$ is $\chi_e \sim (\Delta T)^{-\frac{3}{2}}$ and thus 
 $\gamma_e^0= \frac32.$ Other exponents at $\lambda=0$ can be determined following the scaling relation (\ref{eq:scale}). Thus, the parent 
 universality class  of the electric phase transition is characterized by the exponents,
 \begin{eqnarray}\label{eq:elec_Ising}
\nu_e^0=1;~\gamma_e^0=3/2;~\beta_e^0=1/4;~\delta_e^0=7.
\end{eqnarray}
Using this in Eq. (\ref{eq:exact_ele_AT}) we obtain 
%Continuous variation of exponents as described in Eqs. (\ref{eq:var1}) are (\ref{eq:var1}) are the most generic ones, with 
\be
\label{eq:fg}
 f(\mu) = \frac{\gamma_e^0}{\gamma_e} =\frac{3(4 \mu-3\pi)}{8 \mu - 7 \pi} ;~ g(\mu)= \frac{1} \nu= \frac{ 4 \mu - 3\pi} {2(\mu-\pi)}
\ee
where $\cos\mu = e^{2\lambda} \sinh(2\lambda).$

We perform Monte-Carlo simulations of AT model near the critical self-dual line parametrized by $\lambda,$ \cite{Wegner}
\be \label{eq:TcJclam}
T_{c} =1, \lambda_c=\lambda, J_c = \frac12 \sinh^{-1}(e^{-2\lambda}).
\ee
For a large system ($L=1024$) we calculate the critical exponents $\beta$ and $\gamma$ by varying $T$ and calculating how the order parameters and the susceptibilities vary as a function of $(T_c-T).$ Exponents $\delta_{m,e}$ are obtained from the variations of $M,P$ w.r.t the fields $h,\tilde h.$ To calculate $\beta_{m,e}/\nu$, we employ finite size scaling. Details of the Monte Carlo simulations and resulting critical exponents
for $\lambda=-0.2,-0.1, 0, 0.1,0.2$ are given in the 
Supplemental Material \cite{supp}. The critical exponents compare quite well with the exact values given in Eqs. \eqref{eq:exact_ele_AT} and \eqref{eq:exact_mag_AT}.
This ensures us that the scaling functions obtained in the next section using similar statistical averaging are quite accurate and that, the system size considered here is well within the scaling regime.

\subsection{Invariant scaling functions along the critical line}
Since all critical exponents can change along the line of criticality, we look for the features that remain invariant and ascertain that the super universality hypothesis is in work. Scaling functions are the natural choices. 
\begin{figure}[h]
\vspace*{.1 cm}
\centering
\includegraphics[width=4.25cm]{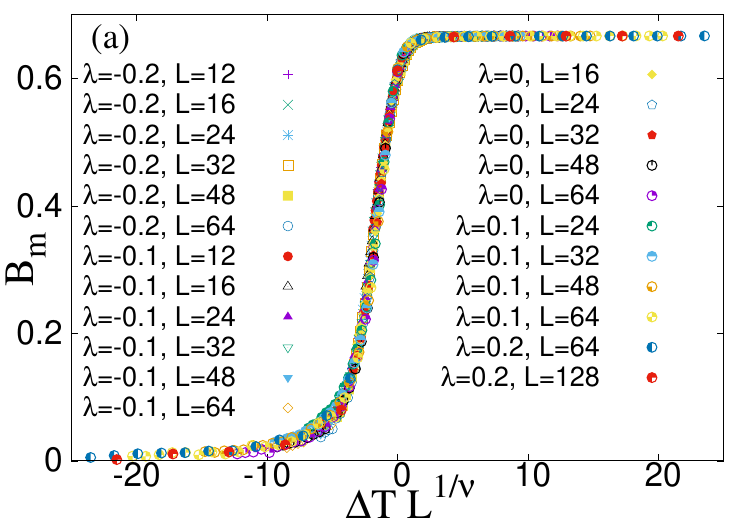}
\includegraphics[width=4.25cm]{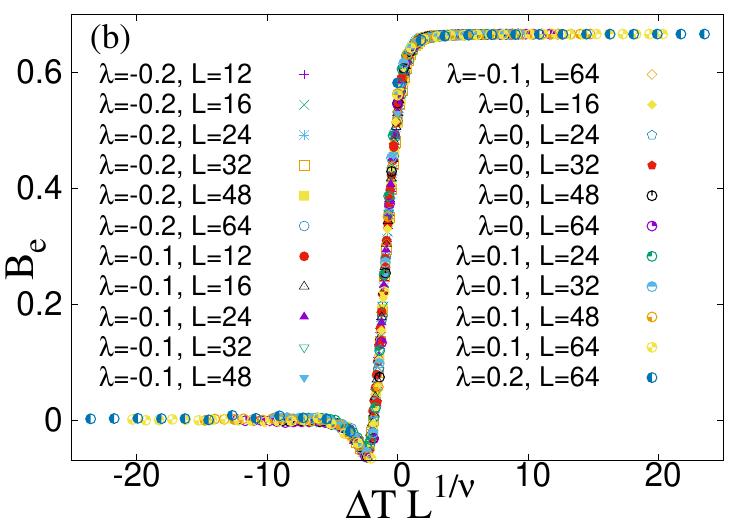}
\includegraphics[width=4.25cm]{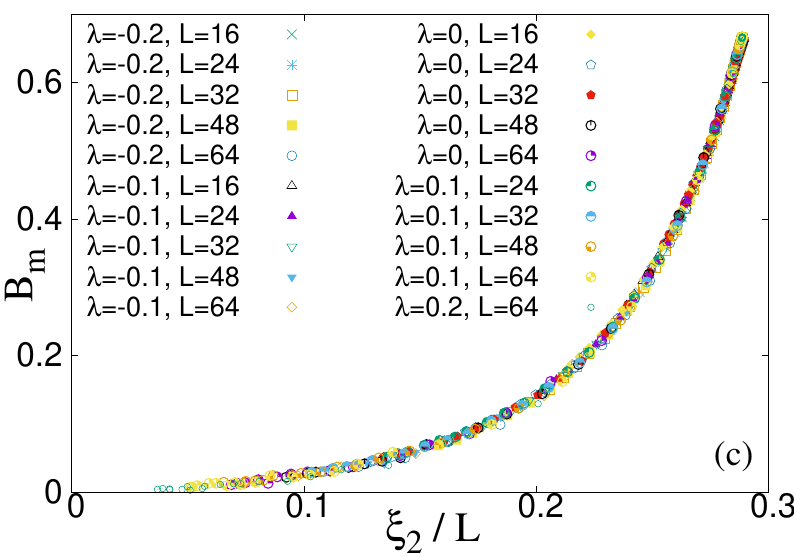}
\includegraphics[width=4.25cm]{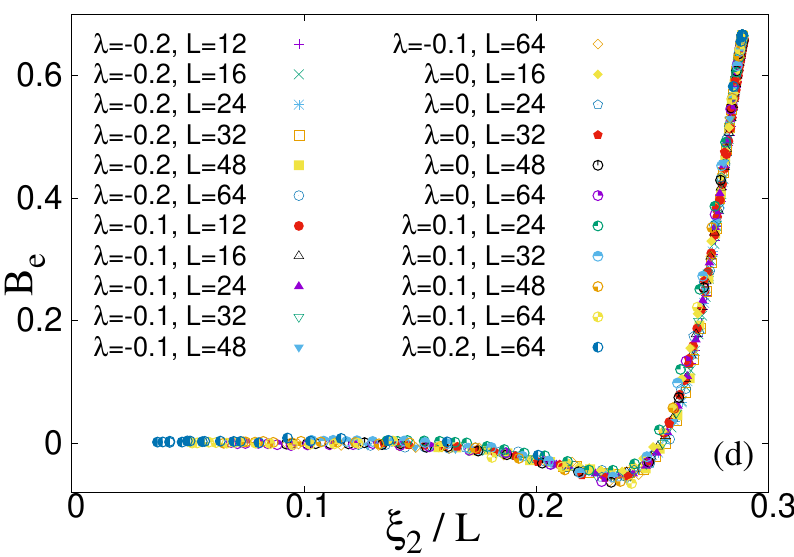}
\caption{(Color online) Binder cumulants $B_{m,e}$ for magnetic, electric transitions. (a) and (b): $B_{m,e}$ vs $\Delta T L^{1/\nu}$. (c)and (d): $B_{m,e}$ vs $\xi_2/L.$ We consider five different  $\lambda$ values $\{-0.2,-0.1,0,0.1,0.2\}$ and  several $L$s for each $\lambda.$ $\xi_2$ is the second-moment correlation length defined in Eq. \eqref{eq:xi2}. In all four cases, the data appear to converge into a unique scaling curve. Data are averaged over $10^7$ samples. Error bars are the same size or smaller than the symbols used.}
\label{fig:B_xi}
\end{figure}

First, we look at the Binder cumulants \cite{K_binder, K_binder1}, which are  RG-invariant.  For the magnetic and the electric transitions they are defined as
\begin{equation}\label{eq:scaling_binder}
B_m= 1- \frac {\langle M^4 \rangle} {3\langle M^2\rangle}; ~~~
B_e= 1- \frac {\langle P^4 \rangle} {3\langle P^2\rangle}
\end{equation}
Since $B_{m,e}$ are dimensionless quantities, they are expected to follow the finite size scaling, 
\begin{equation}
B_m =g_{m}(\Delta T L^{1/\nu}); B_e =g_{e}(\Delta T L^{1/\nu}).
\end{equation}
The system has a dominant correlation length $\xi$ which does not depend on whether one looks at magnetic or electric behaviour and it diverges as $\xi\sim \Delta T^{-\nu}$ at the critical point. In a finite $L\times L$ system $\xi$ is limited by the length $L$ and thus $\Delta T L^{1/\nu}$
can be regarded as  $(\xi/L)^{-1/\nu}.$
 In Fig. \ref{fig:B_xi}(a) and (b) we have shown $B_{m,e}$ as a function of $\Delta T L^{1/\nu}$ obtained for different values of $\lambda$ and system sizes. All the curves appear to collapse into a unique scaling curve.

Another RG-invariant quantity, the second-moment correlation length \cite{VicariRev} provides strong universality checks of the scaling functions. Recently, this idea has been successfully applied in other contexts \cite{Bonati}.
In the AT model, the second-moment correlation length $\xi_2$ is defined as
\be
(\xi_2)^2 = \frac{\sum_{\bf r} r^2 G_{m}({\bf r})} {\sum_{\bf r} G_{m}({\bf r})}, \label{eq:xi2}.
\ee
where $G_{m}({\bf r})$ is the correlation function defined in Eq. \eqref{eq:corr_fn}.
We calculate $\xi_2$ from the Monte Carlo simulations by taking ${\bf r}$ only along $x$- and $y$-directions. 
In Fig. \ref{fig:B_xi}(c) and (d) we have shown $B_{m,e},$ as a function of $\xi_2/L$ obtained for several values of $\lambda$ and system sizes.
All the curves naturally collapse to a single function. Note that the critical behaviour of 
the model at $\lambda= \log(3)/4\simeq 0.2746$ belongs to the universality class of the  Potts model with $q=4$ ($Z_4$ symmetry) and the data has strong finite size correction when $\lambda$ approaches this value. To avoid these ill effects, in Fig. \ref{fig:B_xi}, we consider the data for larger $L$ when $\lambda$ is large. 
\begin{figure}[h]
\vspace*{.1 cm}
\centering
\includegraphics[width=4.25cm]{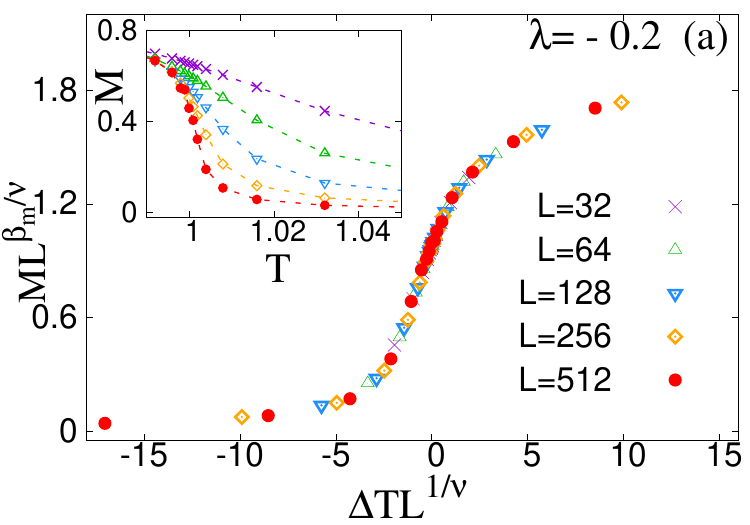}
\includegraphics[width=4.25cm]{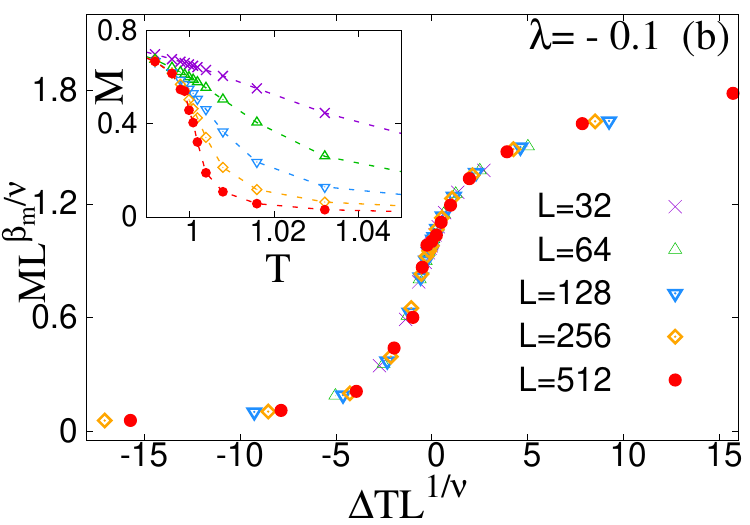}
\includegraphics[width=4.25cm]{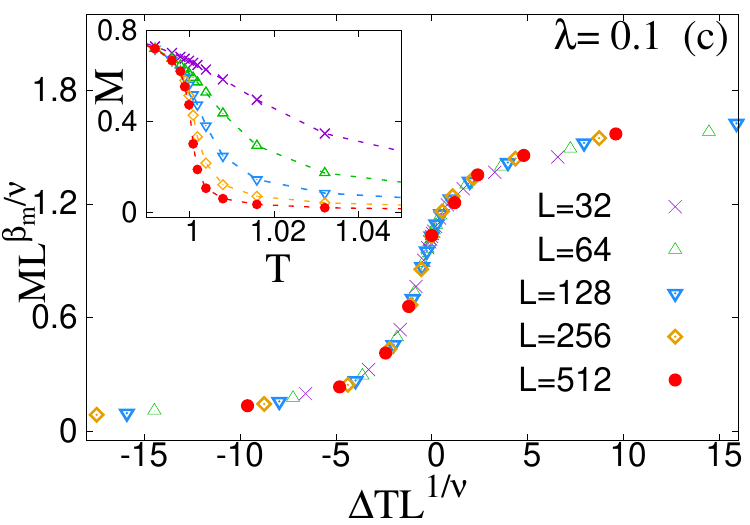}
\includegraphics[width=4.25cm]{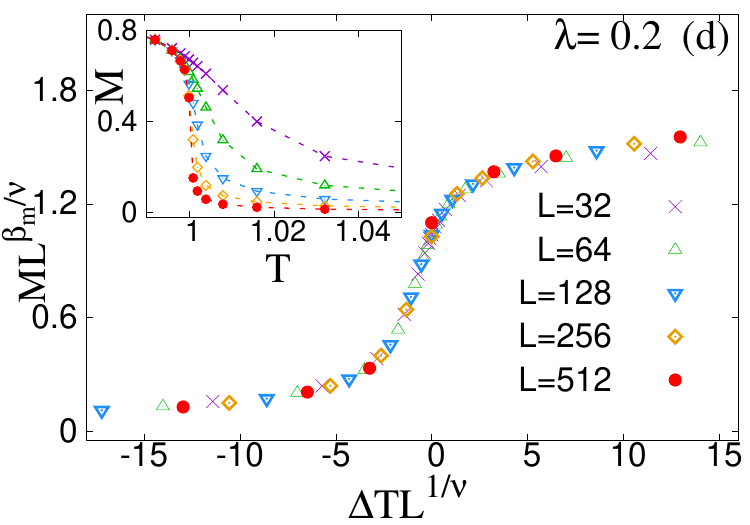}
\includegraphics[width=4.25cm]{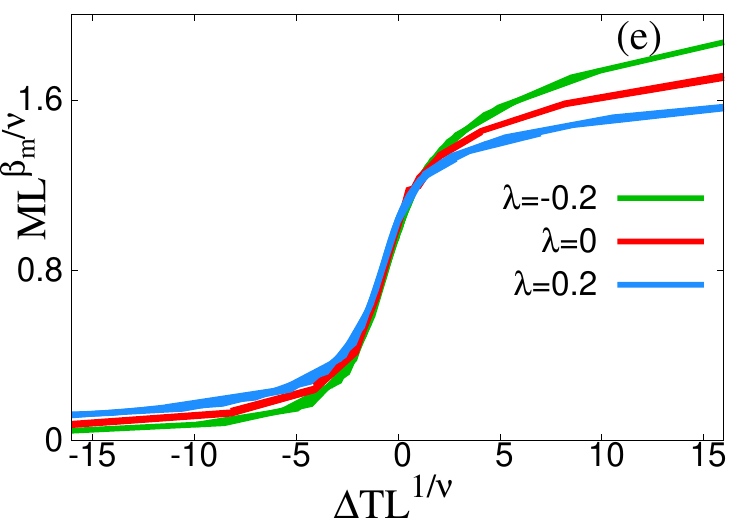}
\includegraphics[width=4.25cm]{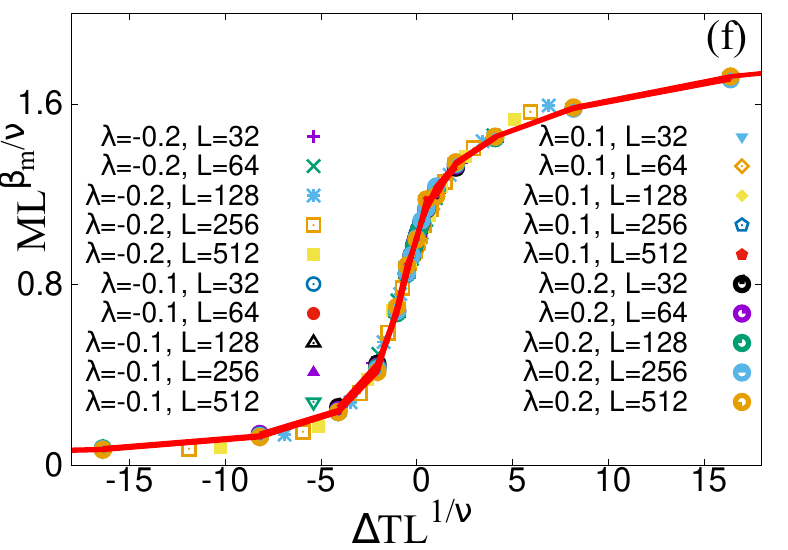}
\caption{(Color online) Data collapses of $ML^{\beta_m/\nu}$ as a function of $\Delta TL^{1/\nu}$ across various system sizes $L=2^5-2^9$ to a unique scaling function observed for (a) $\lambda = -0.2$, (b) $\lambda = -0.1$, (c) $\lambda = 0.1$, (d) $\lambda = 0.2$. In each case inset shows the plots of the order parameter $M$ vs $T$ at corresponding $\lambda.$ The scaling functions obtained for different $\lambda$ are different (shown for $\lambda=-0.2,0,0.2.$ in (e)), but they could be collapsed to a single curve, as shown in (f), by re-scaling the $x$ and $y$ axes.
}
 \label{fig:scaling_M_T}
\end{figure}
\begin{figure}[ht]
\vspace*{.1 cm}
\centering
\includegraphics[width=4.25cm]{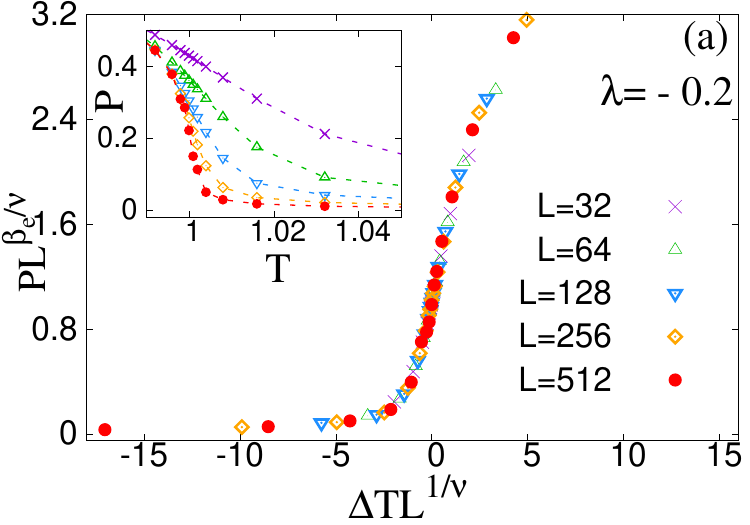}
\includegraphics[width=4.25cm]{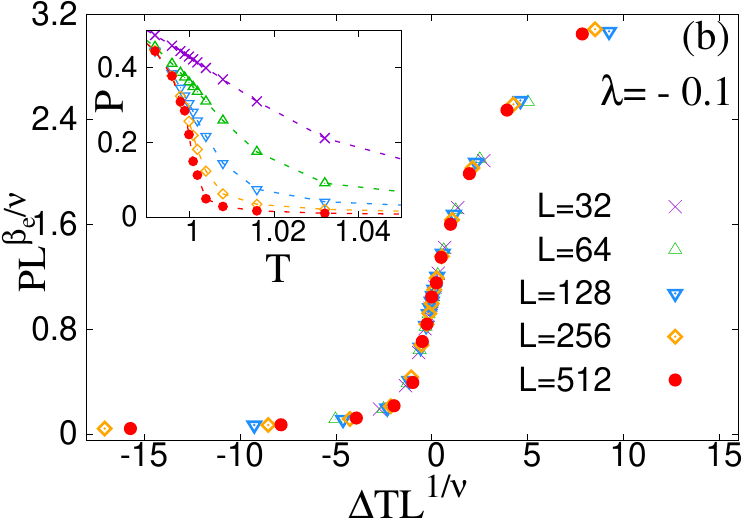}
\includegraphics[width=4.25cm]{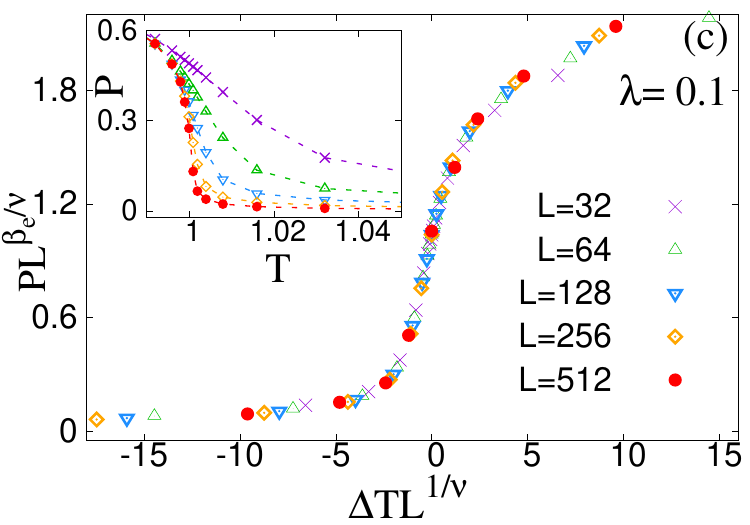}
\includegraphics[width=4.25cm]{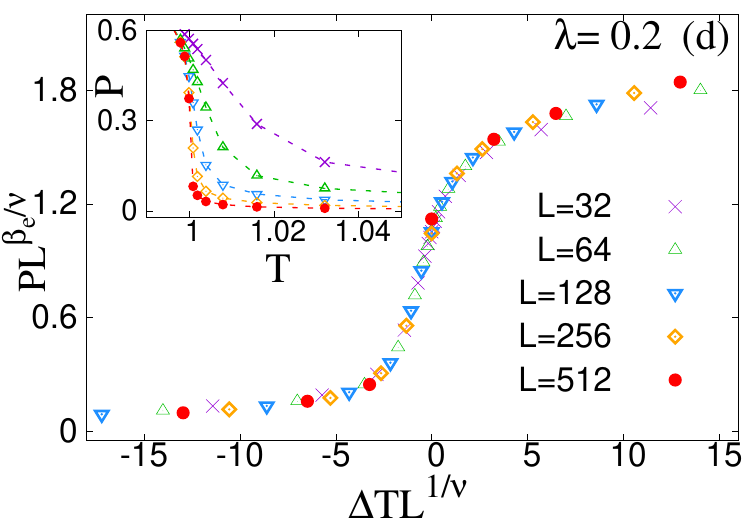}
\includegraphics[width=4.25cm]{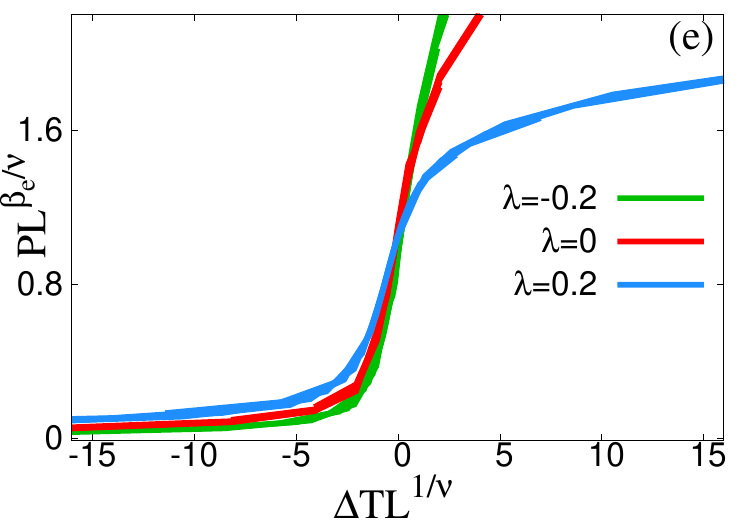}
\includegraphics[width=4.25cm]{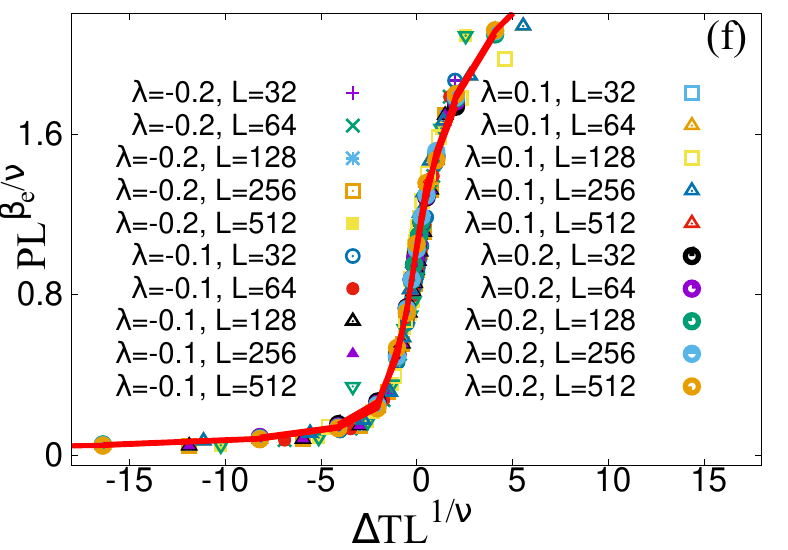}
\caption{(Color online) Data collapses of $PL^{\beta_e/\nu}$ as a function of $\Delta TL^{1/\nu}$ across various system sizes $L=2^5-2^9$ to a unique scaling function observed for (a) $\lambda = -0.2$, (b) $\lambda = -0.1$, (c) $\lambda = 0.1$, (d) $\lambda = 0.2$. In each case inset shows the plots of the order parameter $P$ vs $T$ at corresponding $\lambda.$ (e) Resulting scaling functions are different for different values of  $\lambda=-0.2,0,0.2.$ (f) By re-scaling the $x$ and $y$ axes, we collapse all the scaling functions for $\lambda\ne0$ onto that of $\lambda=0$. Data are averaged over $10^7$ samples or more.
}
 \label{fig:scaling_P_T}
\end{figure}
\begin{figure}[h]
\vspace*{.1 cm}
\centering
\includegraphics[width=4.25cm]{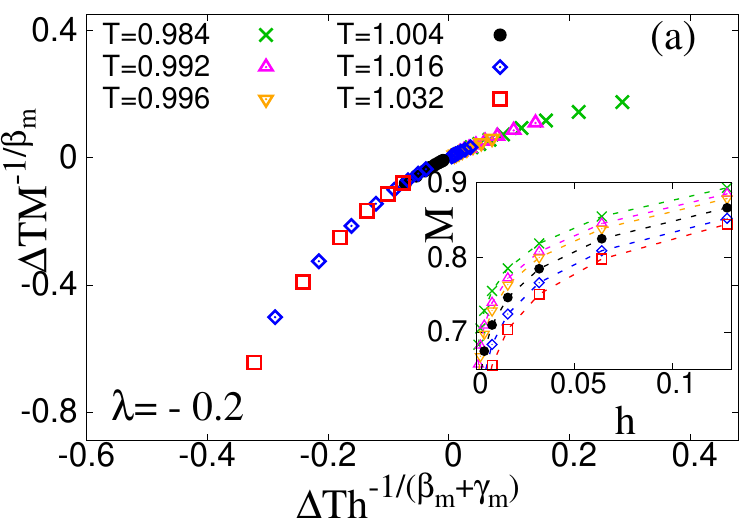}
\includegraphics[width=4.25cm]{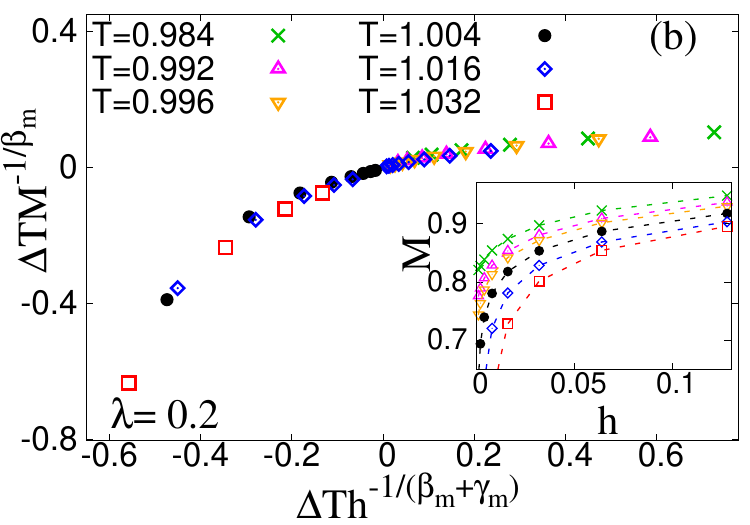}
\includegraphics[width=4.25cm]{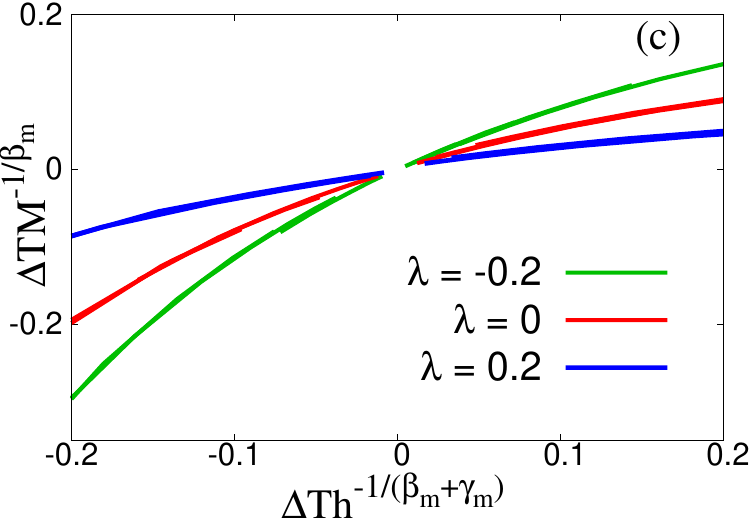}
\includegraphics[width=4.25cm]{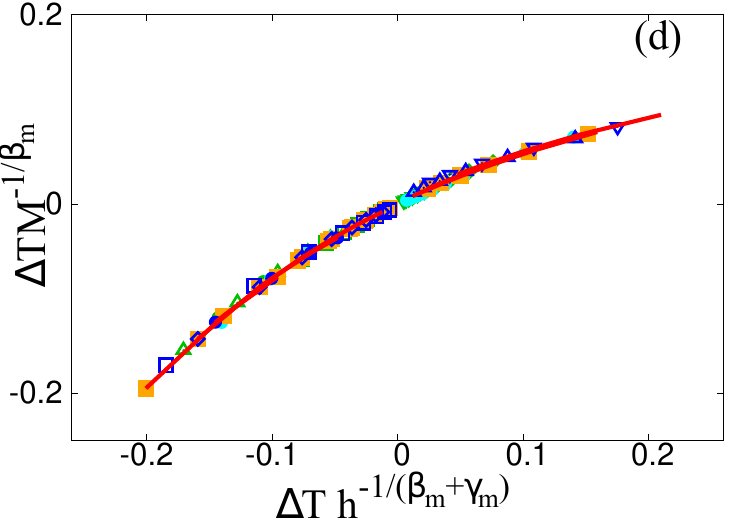}
\caption{(Color online) Data collapses of $\vdt M^{-1/\beta_m}$ as a function of $\vdt h^{-1/(\beta_m+\gamma_m)}$ with different $T$ to a unique scaling functions observed for (a) $\lambda = -0.2$ and (b) $\lambda = 0.2$. In each case, the inset shows the behavior of magnetization $M$ with field $h$ across the temperature $T$ for corresponding $\lambda.$ (c) Scaling functions turn out to be different for different $\lambda.$ (d) By re-scaling $x$ and $y$ axes we collapse all the different  scaling functions (symbols) onto
the scaling function for $\lambda=0$ (red solid line).}
 \label{fig:all_scaling_mag_h}
\end{figure}
\begin{figure}[h]
\vspace*{.1 cm}
\centering
\includegraphics[width=4.25cm]{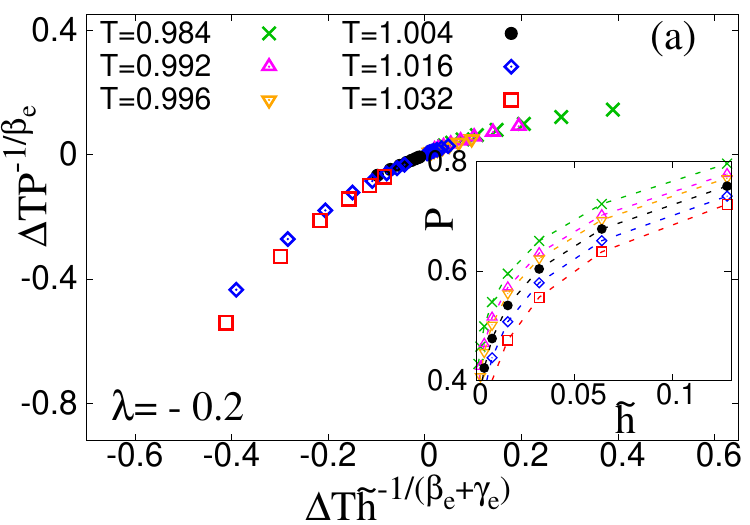}
\includegraphics[width=4.25cm]{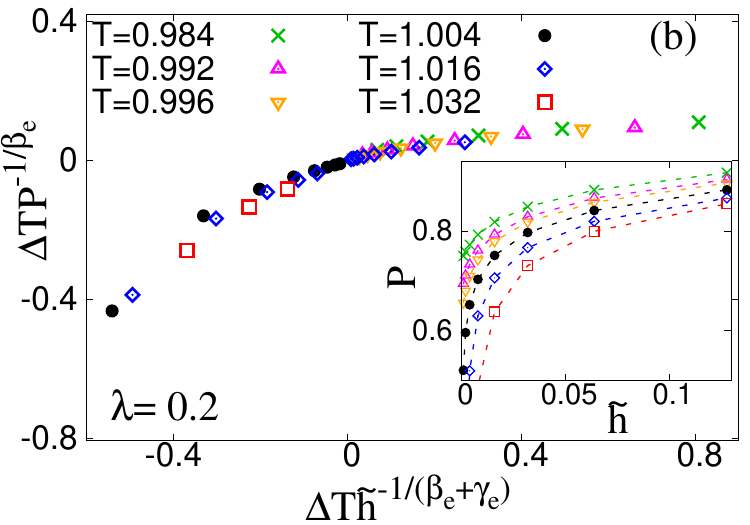}
\includegraphics[width=4.25cm]{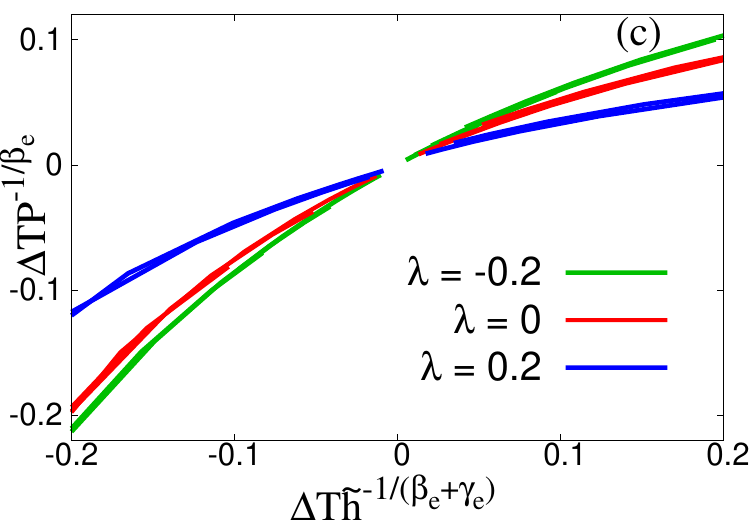}
\includegraphics[width=4.25cm]{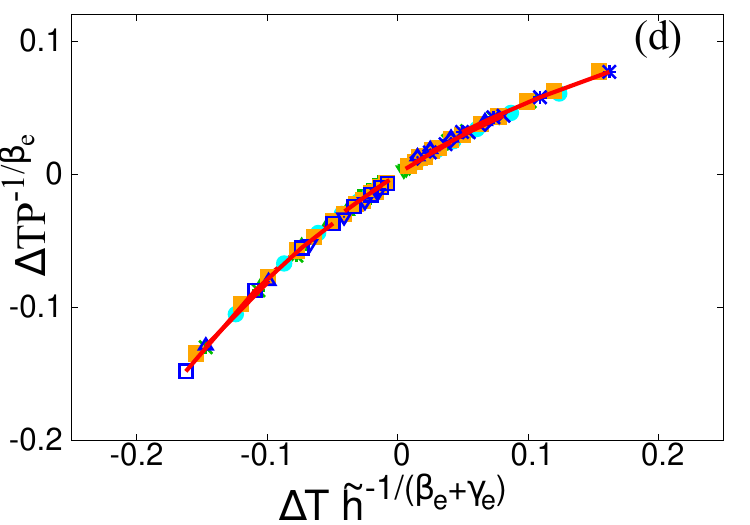}
\caption{(Color online) Data collapses of $\vdt P^{-1/\beta_e}$ as a function of $\vdt \tilde h^{-1/(\beta_e+\gamma_e)}$ with different $T$ to a unique scaling functions observed for (a) $\lambda = -0.2$ and (b) $\lambda = 0.2$. In each case, the inset shows the behavior of polarization $P$ with field $\tilde h$ across the temperature $T$ for corresponding $\lambda.$ (c) Scaling functions turn out to be different for different $\lambda.$ (d) By re-scaling $x$ and $y$ axes we collapse all the scaling functions (symbol) for different $\lambda$ onto the curve corresponding to $\lambda=0$ (red solid line). }
 \label{fig:all_scaling_elec_h}
\end{figure}

The order parameters $\phi_{m,e}\equiv M,P$ also exhibit finite-size scaling, 
\be
\phi_{m,e} = L^{\beta_{m,e}/\nu} f_{m,e}( \Delta TL^{1/\nu}).
\ee
A plot of $ML^{-\beta_m/\nu}$ as a function of $\Delta TL^{1/\nu}$ is shown in Fig. \ref{fig:scaling_M_T}(a)-(d) respectively for $\lambda= -0.2, -0.1, 0.1, 0.2.$ The data for different $L$ collapse to a unique scaling function for each $\lambda.$ However the scaling function for different $\lambda,$ shown in Fig. \ref{fig:scaling_M_T}(e) for $\lambda= -0.2, 0, 0.2$, turns out to be different.
This is because the scaling functions contain non-universal ($\lambda$-dependent) scale factors. Thus, one expects the individual scaling functions of different  $\lambda$  to collapse onto a single curve when 
 the $x$- and $y$-axes are rescaled, which is shown in Fig. \ref{fig:scaling_M_T}(f). 
 A similar data collapse of Polarization, i.e. $PL^{-\beta_e/\nu}$ versus $\Delta TL^{1/\nu}$ for individual $\lambda,$ is shown in Fig. \ref{fig:scaling_P_T}(a)-(d). The individual scaling functions of different $\lambda$ are then made to collapse to a single curve by re-scaling the axes, which is shown in \ref{fig:scaling_P_T}(e)-(f). A good data collapse obtained in both cases indicates that there exists an underlying universal scaling function all along the critical line. Note, that an additional re-scaling of axes is not required for Binder cumulant as, in the thermodynamic limit, $B_{m,e}$ approaches a constant value: $\frac23$ when $T<T_c$ and $0$ when $T>T_c$ \cite{K_binder, K_binder1}.

%figure

Now we turn our attention to field-dependent scaling. 
For a large system the order parameters $\phi_{m,e}$ in the presence of their 
 external field conjugates $B\equiv h,\tilde h$ 
follow a scaling relation \cite{Stanley_1971}, 
\begin{equation} \label{eq:all_scaling}
 \phi_{m,e}=\vdt^{\beta_{m,e}} F_{m,e}(\vdt B^{-1/(\beta_{m,e}+\gamma_{m,e})}).
\end{equation}
For a large system, $L=1024$ we obtain $M$ as a function of $h$ for different values of $T$ near the critical value $T_c=1$ (here $\tilde h=0$). 
A plot of $ \vdt M^{-1/\beta_m}$ as a function of $\vdt h^{-1/(\beta_m+\gamma_m)}$ is shown in Figs. \ref{fig:all_scaling_mag_h} (a),(b) respectively for $\lambda=-0.2,0.2.$ The scaling functions obtained for different $\lambda$ are then compared in Fig. \ref{fig:all_scaling_mag_h}(c). The scaling functions look different, but as expected, they could be collapsed to a unique curve by re-scaling the $x$- and $y$-axes. A similar plot of $ \vdt P^{-1/\beta_e}$ versus $\vdt \tilde h^{-1/(\beta_e+\gamma_e)}$ for $\lambda=-0.2,0.2$ exhibit data collapse in Fig. \ref{fig:all_scaling_elec_h} (a),(b). The individual scaling functions for different $\lambda,$ compared in Fig. \ref{fig:all_scaling_elec_h}(c), are re-scaled to obtain a unique scale function in Fig. \ref{fig:all_scaling_elec_h}(d). 
An excellent match clearly indicates that the scaling properties of both magnetic and electric phase transitions in the AT model can be derived from that of the parent universality class.

\section{Summary} 
Marginal operators, if present in a system, can generate a line of critical points along which the critical exponents may vary continuously. In this article, we introduce a super universality hypothesis (SUH) for the continuous variation of exponents: we propose that, up to constant scale factors, the scaling functions along the critical line must be identical to that of the base universality class even when all the critical exponents vary continuously along the marginal direction. We demonstrate this in the Ashkin Teller model, where the critical exponents of ferromagnetic phase transition vary with the interaction parameter following the weak universality scenario whereas the polarization of the system exhibits a continuous variation of {\it } all  exponents. We calculate several scaling functions and show explicitly that they are indeed universal along the critical line up to certain non-universal scale factors. The scaling functions relating to the Renormalization-group-invariant quantities, like the Binder cumulant $B_{m,e}$, the ratio of the correlation length $\xi\sim (T_c-T)^{-\nu}$ to system size $L,$ and the ratio of the second moment correlation length $\xi_2$ to $L$ do 
 not require any additional scaling - they naturally remain invariant along the critical line. 
However, the hyperscaling relations between the critical exponents are obeyed as long as the system has a unique diverging length scale.

The SUH is helpful in identifying whether two sets of critical exponents, both satisfying hyperscaling relations, belong to different universality classes or are only instances of a super universality class generated by a marginal operator. It suggests that if the change of exponents is caused by a marginal operator, the underlying scaling functions must be unique up to multiplicative scale factors. 
This question is quite relevant in the study of phase transition in experimental systems, where there is only a limited set of measured exponents; if they differ, more often than not, their critical behavior is assigned to different universality classes.

In our opinion, the super universality hypothesis is quite general and it can be applied to other systems where a marginal parameter leads to continuous variation of critical exponents. While the validity of hyperscaling relations provides a guideline on the functional form of the continuous variation, SUH suggests that when all critical exponents vary along a critical line, the invariant scaling functions are the ones that carry forward the universal features of the parent universality class.

{\it \bf Acknowledgement:} The authors thank the anonymous referees for their critical comments and constructive suggestions that helped us improve the quality of the manuscript immensely. IM acknowledges the support of the Council of Scientific and Industrial Research, India in the form of a research fellowship (Grant No. 09/921(0335)/2019-EMR-I).

\newpage
\setcounter{equation}{0}
\setcounter{figure}{0}

\renewcommand{\theequation}{S\arabic{equation}}
\renewcommand{\thefigure}{S\arabic{figure}}
\renewcommand{\bibnumfmt}[1]{[S#1]}
\renewcommand{\citenumfont}[1]{S#1}

\onecolumngrid
\clearpage
\begin{center}
{\Large{\bf Supplemental Material for ``Hidden Super Universality in Systems with Continuous Variation of Critical Exponents"}}
\end{center}
\vspace{0.5cm}
\begin{center}
In this supplemental material, we calculate the critical exponents of the magnetic and electric phase transitions from Monte Carlo simulations of the Ashkin Teller model.
\end{center}

\begin{figure*}[h]
%\vspace*{.5 cm}
\centering
\includegraphics[width=4.25cm]{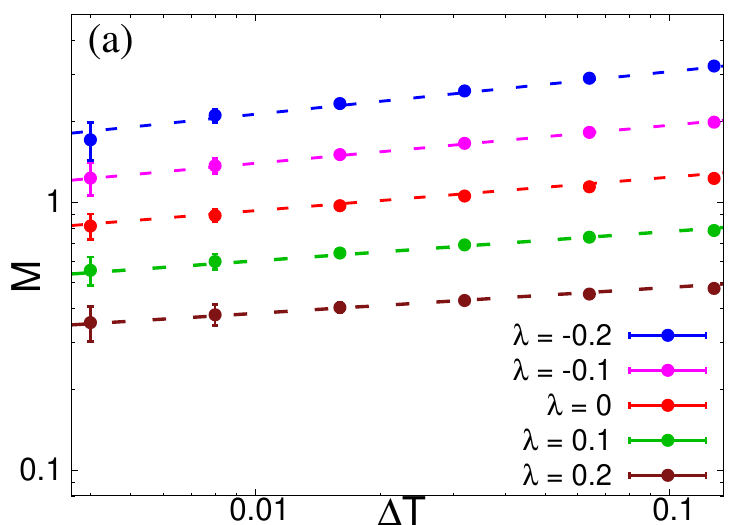}%\hspace{.2 cm}
\includegraphics[width=4.25cm]{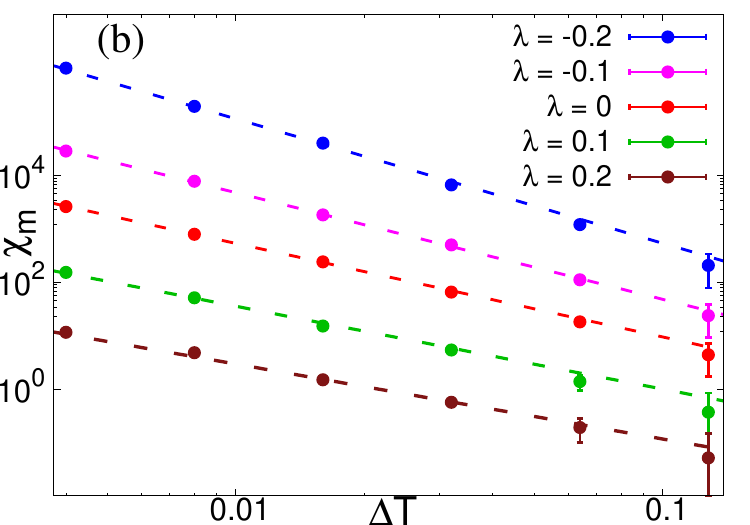}
\includegraphics[width=4.25cm]{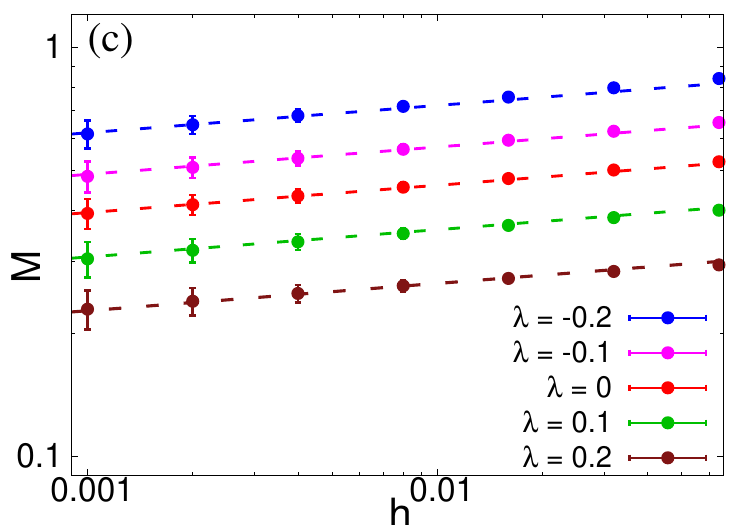}%\hspace{.2 cm}
\includegraphics[width=4.25cm]{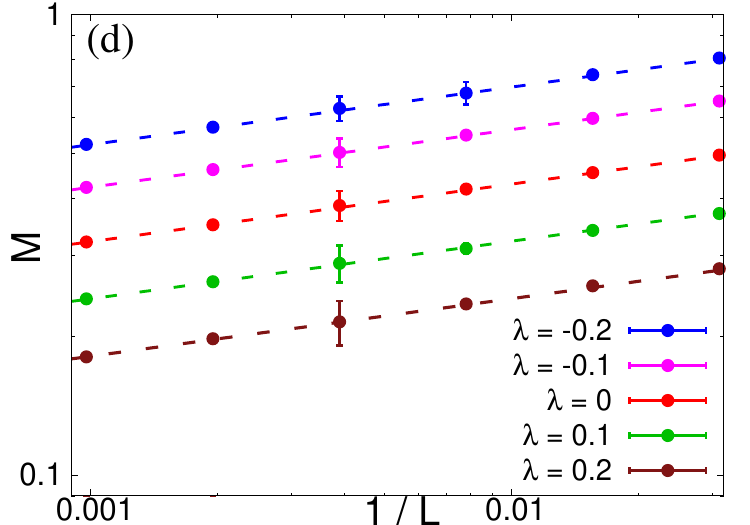}
\caption{
Critical exponents of magnetic transition from log-scale plots: (a) $\beta_m$ from $M$ vs. $\Delta T,$ (b) $\gamma_m$ from $\chi_m$ vs. $\Delta T,$ (c) $\delta_m$ from $M$ vs. $h$ at the $T=T_c,$ and (d) $\frac{\beta_m}{\nu}$ from $M$ vs. $L^{-1}.$ Dashed lines with slopes the same as exactly known exponent values are drawn for comparison.In (a)-(c), $L=1024.$ In all cases, data is averaged over $10^6$ or more samples, and the $y$-axis is scaled by arbitrary factors for better visibility. 
\ref{tab:cri_exponents}}
 \label{fig:exponents_mag}
\end{figure*}

\begin{figure*}[h]
%\vspace*{.5 cm}
\centering
\centering
\includegraphics[width=4.25cm]{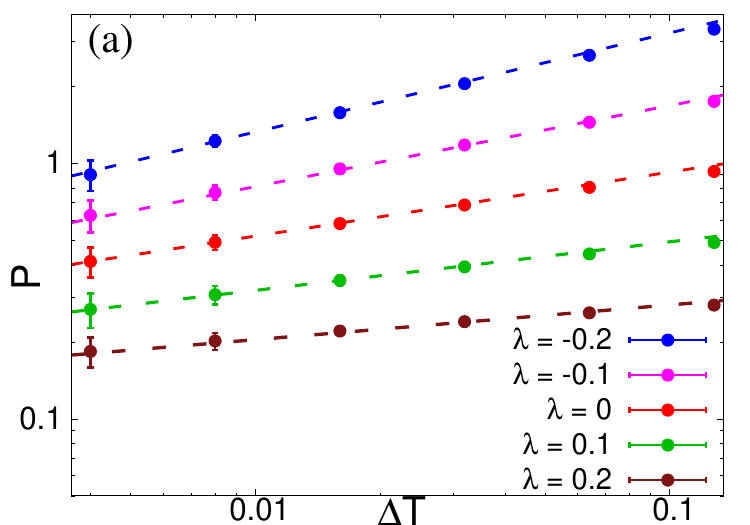} %\hspace{.2 cm}
\includegraphics[width=4.25cm]{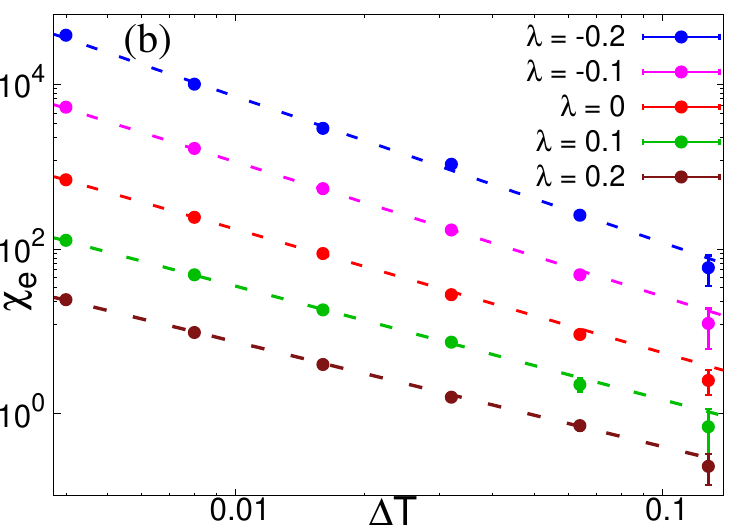}
\includegraphics[width=4.25cm]{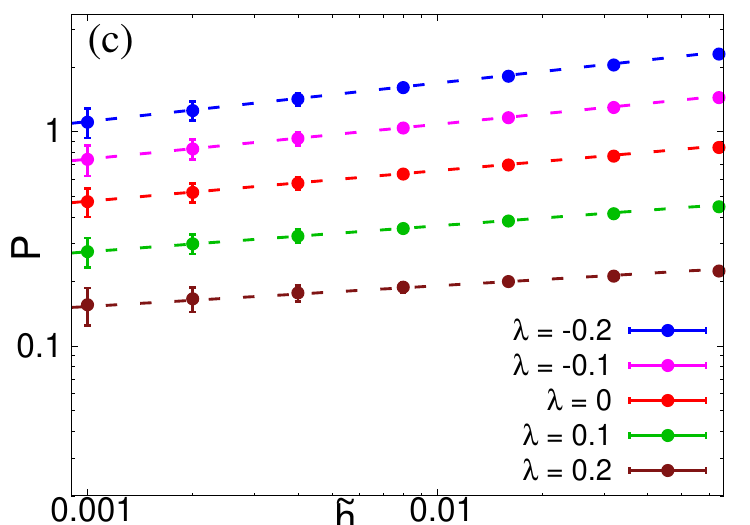} %\hspace{.2 cm}
\includegraphics[width=4.25cm]{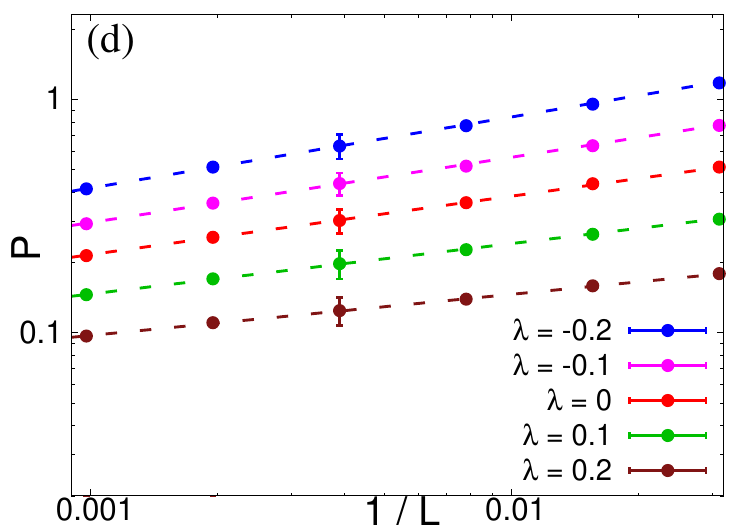}
\caption{ Critical exponents of electric transitions, obtained from of $P$ and $\chi_e.$  The parameters are identical to that of Fig. \ref{fig:exponents_mag}. }
 \label{fig:exponents_elec}
\end{figure*}

\vspace{0.5cm}
The critical exponents of magnetic and electric phase transitions occurring in the Ashkin Teller (AT) model are known exactly and listed in Eqs. (9) and (10) of the main text. Here we calculate them from Monte Carlo simulations of the model and benchmark it against the known exact results. We set the parameters near the critical self-dual line (following Ref. [49] of the main text) parameterized by the interaction parameter $\lambda,$
\be \label{eq:TcJclam}
T_{c} =1, \lambda_c=\lambda, J_c = \frac12 \sinh^{-1}(e^{-2\lambda}).
\ee
To estimate the exponents $\beta_{m,e}$ we calculate $M, P$ and their respective variances $\chi_{m,e}$ for different temperatures $T=T_c-\Delta T$ and plot them as a function of $\Delta T,$ in log-scale.
For these simulations, we use a large system ($L=1024$), and statistical averaging is done over $10^7$ samples or more samples. The simulation is repeated for $\lambda =$ $-0.2$, $-0.1$, $0$ (Ising), $0.1$ and $0.2$ 
The results (in symbol) are then compared with the exact values (dashed-line) known for corresponding $\lambda.$ Figures \ref{fig:exponents_mag}(a),(b) shows the plots $M$ 
and $\chi_m,$ whereas Figs. \ref{fig:exponents_elec}(a),(b) corresponds to $P$ and $\chi_e$ respectively. The field exponents $\delta_{m,e}$ are obtained from $M,P$ as a function of $h,\tilde h,$ at $T=T_c=1,$ where 
$\tilde h$ and $h$ in Eq. \eqref{eq:AT_H_h} of the main text are set to zero respectively. The log-scale plots of $M,P$ versus $h,\tilde h$ for different $\lambda$ are 
shown Figs. \ref{fig:exponents_mag}(c), \ref{fig:exponents_elec}(c).

\begin{figure*}[h]
%\vspace*{-3.1 cm}
\centering
\includegraphics[width=4.25cm]{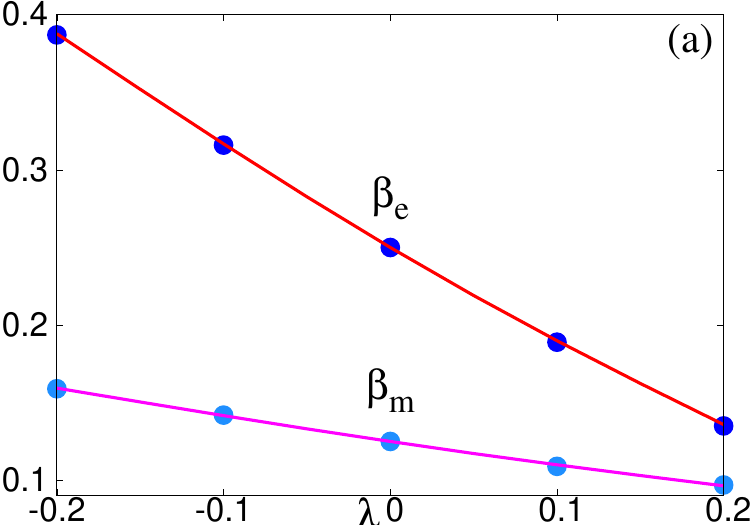}
\includegraphics[width=4.25cm]{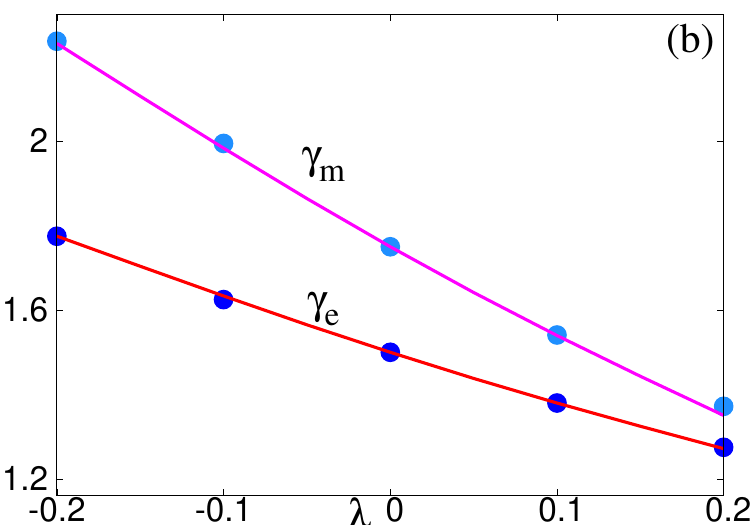}
\includegraphics[width=4.25cm]{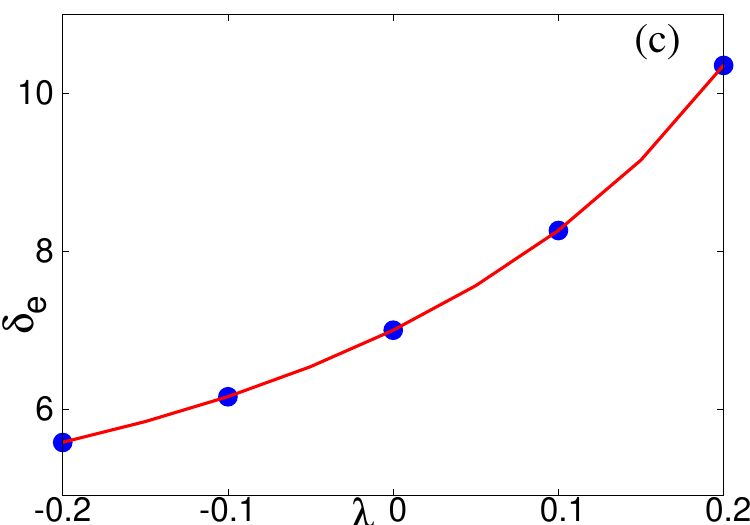}
\includegraphics[width=4.25cm]{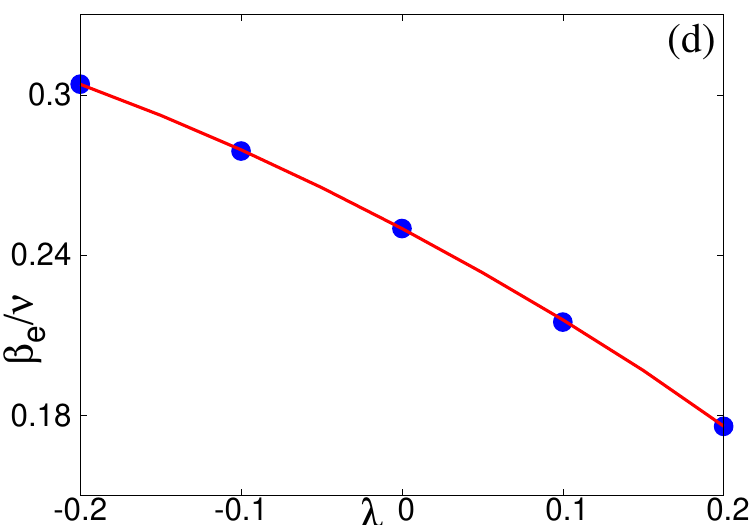}
\caption{ Critical exponents obtained from Monte Carlo simulations (symbols) are compared with their exact theoretical values given in Eqs. (9) and (10) of the main text (solid lines). }
%The functional forms of the magnetic and electric exponents provided by SUH in Eqs. (15) and (18) (main text) are also consistent with the corresponding solid lines obtained from their exact values. (a) $\beta_{m,e}$ with $\lambda$, (b) $\gamma_{m,e}$ with $\lambda$, (c) $\delta_{e}$ with $\lambda$, and (d) $\beta_{e}/\nu$ with $\lambda$}
\label{fig:exponents_lambda}
\end{figure*}

We also calculate the exponents $\beta_{m,e}/\nu.$ We set $T=T_c$ and compute $M,P$ for systems of different size $L.$ For finite systems the correlation length $\xi$ is limited by $L$ and thus $M\sim L^{-\beta_m/nu},$ and $P\sim L^{-\beta_e/nu}.$ The log scale plots of $M,P$ versus $1/L,$ for different $\lambda$ result in straight lines of slope $ \beta_{m,e}/\nu.$ Our estimates of the critical exponents are listed in Table \ref{tab:cri_exponents}; their variation as a function of $\lambda$ (symbol) are compared with the exact functional form (solid lines), in Figs. 
\ref{fig:exponents_lambda} (a)-(d).

\newpage
\begin{table*}[h]
\caption{\label{tab:cri_exponents} Critical exponents of magnetic and electric phase transitions in Ashkin Teller model, obtained from Monte Carlo simulations for different interaction parameter $\lambda.$}
\begin{ruledtabular}
\begin{tabular}{ccccccccc}
 \textrm{$\lambda$}&
\textrm{$\beta_{m}$}&
\textrm{$\gamma_{m}$}&
\textrm{$\delta_{m}$}&
\textrm{$\frac{\beta_m}{\nu}$}&
\textrm{$\beta_{e}$}&
\textrm{$\gamma_{e}$}&
\textrm{$\delta_{e}$}&
\textrm{$\frac{\beta_e}{\nu}$}\\ \colrule
-0.2 & 0.159(6) &	2.237(1) & 15.000(7) & 0.125(2) & 0.387(5) & 1.775(8) &	5.578(9) & 0.304(8)\\
-0.1 & 0.142(3) & 1.995(7) & 15.000(1) & 0.125(0) & 0.316(9) &	1.634(3) &	6.156(2) & 0.278(7)\\
0 &	0.125(1) & 1.750(8) & 15.000(1) & 0.125(0) & 0.250(7)	& 1.501(2)	& 7.000(8) & 0.250(0)\\
0.1& 0.109(7) & 1.541(9) & 15.000(2) & 0.125(2) & 	0.189(6) & 1.379(2) & 8.264(4) & 0.216(7)\\
0.2 & 0.097(4) & 1.372(1) & 15.001(8) & 0.125(1) & 0.135(1) & 1.272(6) & 10.359(5) & 0.176(2)\\
\end{tabular}
\end{ruledtabular}
\end{table*}

\end{document}